# A Micro Architectural Events Aware Real-Time Embedded System Fault Injector

Enrico Magliano, Alessio Carpegna, Alessandro Savino, Stefano Di Carlo
*Department of Control and Computer Engineering(DAUIN)*
Politecnico di Torino
Turin, Italy
{enrico.magliano, alessio.carpegna, alessandro.savino, stefano.dicarlo}@polito.it

*Abstract*—In contemporary times, the increasing complexity of the system poses significant challenges to the reliability, trustworthiness, and security of the Safety-Critical Real-Time Embedded Systems (SACRES). Key issues include the susceptibility to phenomena such as instantaneous voltage spikes, electromagnetic interference, neutron strikes, and out-of-range temperatures. These factors can induce switch state changes in transistors, resulting in bit-flipping, soft errors, and transient corruption of stored data in memory. The occurrence of soft errors, in turn, may lead to system faults that can propel the system into a hazardous state. Particularly in critical sectors like automotive, avionics, or aerospace, such malfunctions can have real-world implications, potentially causing harm to individuals.

This paper introduces a fault injector designed with the novelty to facilitate the monitoring, aggregation, and examination of micro-architectural events. This is achieved by harnessing the microprocessor's Performance Monitoring Unit (PMU) and the debugging interface, explicitly focusing on ensuring the repeatability of fault injections. The fault injection methodology targets bit-flipping within the memory system, affecting CPU registers and RAM. The outcomes of these fault injections enable a thorough analysis of the impact of soft errors in the final output and timing predictability demanded by SACRES.

## I. INTRODUCTION

The evolution of embedded systems has resulted in increased performance to meet the demands of progressively sophisticated applications. However, this progress introduces greater system complexity, employing multi-core systems with more Central Processing Units (CPUs) and complex memory hierarchies with various levels of caches. Consequently, modern chips become more susceptible to soft errors, particularly when combined with a stratified software stack, propagating faults with potentially tragic consequences for application execution [1]. Environmental phenomena interacting with the circuit can induce bit-flipping and subsequent data loss, resulting in soft errors [2].

Numerous studies focus on analyzing and mitigating this issue to ensure the reliability and trustworthiness of Safety-Critical Real-Time Embedded Systems (SACRES). In this context, Fault Injectors (FIs) are fundamental tools capable of simulating the occurrence of soft errors or injecting faults into the system [3]–[5]. Various possibilities exist for implementing a FI. For instance, hardware-based implementations, such as those presented in [6]–[8], are accurate but exhibit low controllability. Conversely, Software-Implemented FI, like those in [9]–[12], generally incur overhead in terms of time but offer higher controllability and observability.

This work presents a new FI environment tailored for SACRES capable of injecting bit-flips in real embedded hardware boards exploiting the debug unit of modern CPUs. It guarantees complete controllability with a low overhead in terms of time. Despite its portability on several embedded platforms, the FI has been customized and tested to work with the Xilinx Zynq® boards running the FreeRTOS [13] embedded Operating System (OS). This represents a typical setup used in several real embedded applications. Historically, works on reliability analysis in embedded systems mainly focused on the application layer [14], [15]. Only limited studies emphasize the importance of the OS in reliability analysis in embedded systems [16]. An initial effort to analyze faults in OS space has been proposed in [17], and more recent works, like [18]–[20], target specific OSs, covering most OS data structures. Due to this reason, the proposed FI is designed to be as general as possible, capable of injecting faults in CPU registers and RAM, encompassing both OS and application space, at any time from OS bootstrapping to the end of the computation, these features ensure complete controllability of the injection process.

Previous work demonstrates the possibility of realizing a FI using the GNU Debugger in embedded systems [12], [21]. However, when dealing with real-time systems, the correctness of the computation is not the only important parameter. Predictability in the execution is an essential requirement. Therefore, a crucial aspect of reliability analysis of real-time embedded systems involves monitoring architectural events characterizing system execution, potentially indicating if a soft error occurs [22] or if time constraints are not respected. In modern CPUs, the Performance Monitoring Unit (PMU) tracks these events, including clock cycles, writing and reading operations, cache hits and misses, and branch statistics. To consider this aspect, the proposed FI is designed to exploit the PMU to profile desired architectural events using the Hardware Performance Counters (HPCs) and collect them at the end of the computation. Apart from performing reliability analysis, the capability of profiling hardware events in the presence of faults has demonstrated interesting potential to build efficient machine learning-based fault detectors [22], [23].

Eventually, the ability to inject faults into real hardware boards guarantees high throughput and experiments performed in a highly realistic setup.

The paper is structured as follows: Section II outlines the methodologies employed to implement the FI. Section III details and discusses the results of utilizing our FI. Finally, Section IV provides a comprehensive summary of this study.

## II. METHODOLOGY

This section outlines the proposed FI architecture and main design characteristics.

### A. High-level architecture

Figure 1 shows the high-level architecture of the FI framework.

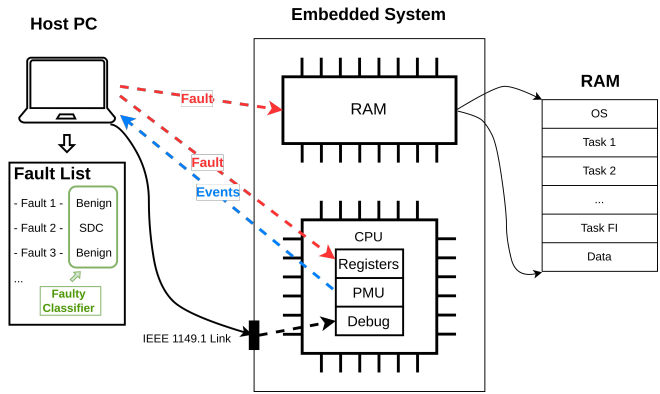

Fig. 1: *High-level architecture of the FI framework. A host machine interacts with the target embedded system through an IEEE 1149 Link to orchestrate the injection experiments.*

Modern microprocessors have a dedicated debug unit, offering precise control over software execution. This unit can temporarily halt the application execution, read/write CPU registers and memory, and resume normal computation. The computation can be stopped at the assembly instruction level by inserting hardware breakpoints, providing excellent temporal control. The CPU debugging unit, controllable via an external host using the IEEE 1149.1 JTAG protocol [24], introduces minimal time overhead when injecting faults into an embedded application.

The FI initiates an experimental campaign based on a list of target faults; each represented as a tuple $(target\_structure, target\_instruction)$. The $target\_structure$ denotes the fault location, encompassing CPU registers and memory locations storing code and data. Conversely, the $target\_instruction$ represents the instant when the fault is injected, using breakpoints to halt the execution. While this approach is straightforward, it may reduce time granularity in looped code (targeting only the first iteration). Alternative approaches, such as timers on the host computer, can overcome this limitation. This tuple guarantees full controllability since one tuple is directly connected to a unique injection location and moment during the target software execution.

Typical configurations in the target embedded system setup involve an embedded operating system (e.g., FreeRTOS) running predefined tasks, each utilizing memory. A Real-Time scheduler handles task scheduling. Embedded real-time operating systems usually lack memory protection mechanisms, making them susceptible to corruption. The FI can inject faults in real-time, including OS bootstrap and task execution. The implemented Single Bit Upset (SBU) fault model can be expanded to Multi Bit Upsets (MBUs). After each injection experiment, the FI classifies the faulty execution outcome by comparing a fault-free *golden execution* with the *faulty execution*. Possible fault classifications are: (i) *Benign* (outputs are equal), (ii) *Silent Data Corruption (SDC)* (different outputs), or (iii) *Other* (e.g., crashes or hangs due to reasons like pointer corruption) [25].

Eventually, the FI enables the profiling of hardware events during injection experiments, necessitating instrumentation with a dedicated injection task. The debug-based fault injection approach ensures full repeatability of faulty executions, which is crucial as most CPU architectures have few HPCs despite numerous trackable architectural events. This limitation means that, in a regular execution, only a small subset of events can be monitored. The ability to repeat injection experiments with reproducible results allows for tracking different events across various executions. Leveraging the available HPCs on the CPU's PMU, the FI repeats faulty executions for each generated fault, collecting all desired architectural events.

### B. Implementation details

Without losing in generality, this section provides additional implementation details targeting Xilinx Zynq® boards running the FreeRTOS [13] embedded OS.

The FI architecture comprises two distinct executable modules: one runs on the embedded system (*target*), and the second (*host*) operates on the host machine (see Figure 1). The host, implemented in Python, exploits the JTAG [24] protocol to govern the debugging of the running application using specific architectural commands, enabling control of the target application execution through debug commands. More in detail, the host resorts to the Xilinx Software Command-Line Tool (XSCT) [26], which serves as an interactive and scriptable command-line interface to Xilinx SDK. Built on top of the Tools Command Language (Tcl), XSCT supports various actions, such as creating and configuring hardware, board support packages, application projects, and flashing boot images. The Pexpect module is employed to script these commands, enabling Python to spawn child applications (XSCT console), control them, and respond to expected patterns in their output.

The target represents the complete embedded software operating on the embedded board. It includes the OS and the application tasks compiled in a single Executable and Linkable Format (ELF) binary file and then flashed on the target hardware.

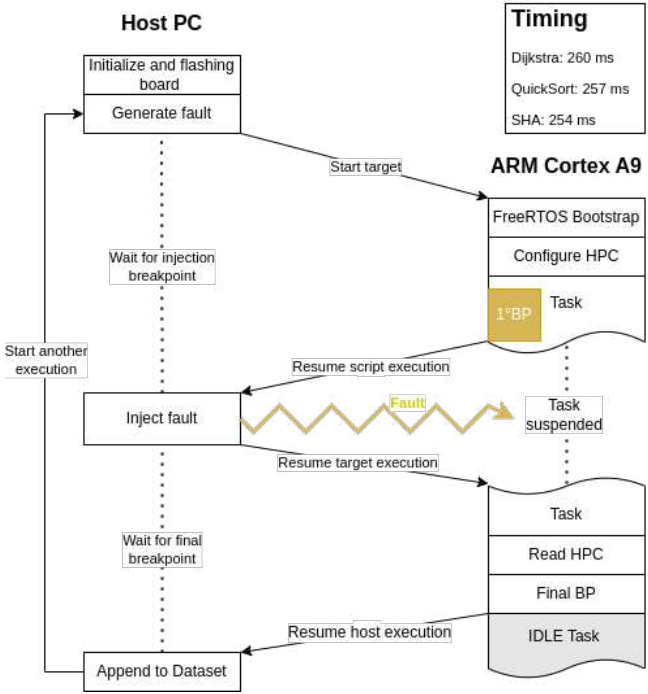

Fig. 2: *Flow software execution over the host that runs the FI script and the target embedded system on which is running the real-time OS and the benchmark task, and in which way they interact with a time reference for one loop in ms of each benchmarks.*

The interaction between the two modules is shown in Figure 2. The host initializes the injection process by setting up the board and generating a fault list, randomly selecting tuples of locations and timing as described in Section II-A. The FI target includes a CPU register/memory address and bit position, while timing refers to a memory address in the code space for setting a breakpoint. A final breakpoint is set after the target tasks completion, crucial for reading the HPCs in the PMU. This ensures that the real-time OS (e.g., FreeRTOS) does not persist in executing the IDLE task, hindering board execution completion. The host then enters the injection loop, setting breakpoints for each tuple and bootstrapping the board. It waits until the board completes the initial execution phase, covering bootstrapping, HPC configuration, and task initiation.

As soon as the injection breakpoint is reached, the host regains the control to inject the faults, remove the breakpoint, and resume execution. Without any further breakpoint, the application tasks are free to run toward the final breakpoint, signaling the host about the end of execution. The host is programmed to wait for a timeout exception if something prevents such an end. To classify the execution, the host verifies the program's output to mark the execution as a Benign or SDC based on the expected result, or if a timeout occurs, the fault is marked as crashes/hangs (Other). Since the FI has to comprise different FI targets, the host repeats the described flow for all the faults.

The target software requires instrumentation to support fault injection, including fault injection operations and event tracking via `Tcl` scripts. Listing 1 presents the pseudo-code for the main instrumentation of the real-time OS. It outlines a common main for the OS, defining a specific task for fault injection (line 2) and starting the scheduler (line 3). Since injection uses the debugger, application tasks remain unaltered, with only the necessary operations wrapped around them for proper interaction with the PMU: initialize PMU before target tasks (line 7) and read HPCs after task completion (line 9). PMU configuration utilizes specific assembly instructions, such as `MCR <register> <value>` in ARM architecture. The procedure is analogous to reading from PMU at the end of the benchmark, but instead of using the MCR instruction, the `MRC <register> <value>` instruction is utilized.

Listing 1: *Task Creation Pseudo Code over the target embedded system*

```
int OS_main (  ){
    taskCreate( faultInjectorTask );
    taskStartScheduler();
    for( ;; ); // endless loop for the OS
}
static void faultInjectorTask( ){
    confPMU();
    startTasks();
    readPMU();
    // taskDelete( NULL );
} // <- final breakpoint point here
```

Ultimately, FI ensures the reproducibility of each injection, a crucial aspect when gathering architectural events that often surpass the available performance counters. This is achieved by running FI multiple times for each fault, altering the tracked architectural events. The process involves a list of selected architectural events, and based on the number of these events and available HPCs, FI repeats the injection with the same parameters while changing the events to be collected. However, this feature introduces a performance impact, multiplying the executions needed to complete the fault injection campaign, as defined as # repetitions = # of events/# of HPC.

## III. RESULTS

This section showcases the capabilities of the proposed FI on a selected experimental setup.

The target architecture is a Dual-core ARM Cortex A9 processor on the Xilinx PYNQ Z2 board. The PYNQ Z2, an open-source project from Xilinx [27], is based on the Xilinx Zynq SoC. The Arm Cortex A9, a 32-bit dual-core processor widely adopted platform, is based on `ARMv7` specifications, supports the Thumb and Thumb-2 instruction sets, and incorporates coherent cache management. It works at 650 MHz, and its features provide a cost-effective and performance-efficient representation of real-world environments. Moreover, the board features a USB port for configuration through JTAG.

The processor incorporates a configurable PMU designed for easy customization to monitor a diverse set of 168 architectural events. In the case of the ARM Cortex-A9, its PMU provides six HPCs, each associated with event-type registers that specify the tracked event, along with additional configuration registers. Access to these registers is facilitated through the internal CP15 interface, as elaborated in the ARM Architectural Reference Manual [28].

The application setup includes the FreeRTOS[13] operating system running a set of benchmarking tasks taken from the MiBench [29] suite. FreeRTOS is a widely-used open-source Real-Time Operating System (RTOS) designed for microcontrollers and small microprocessors. With a focus on precise timing and responsiveness, FreeRTOS is a popular choice in critical applications such as aerospace and industrial automation. Its key features include task scheduling and inter-task communication, making it valuable for developing SACRES. Written primarily in C, FreeRTOS is known for its portability and flexibility across various hardware platforms, contributing to its broad adoption in the embedded systems domain. As benchmarks, this work uses a subset of the embedded benchmarks (`QSort`, `SHA`, `Dijkstra`). They are selected due to the varying characteristics in terms of time and complexity: (i) computationally intense tasks (`SHA` and `Dijkstra`), and (ii) memory-intense tasks (`QSort`). This diversity allows for a more detailed analysis of the effect of the fault injection when the execution time is affected.

A total of 9 fault injection campaigns were conducted (see Figure 4) with the three target benchmarks (`Dijkstra`, `QuickSort`, `SHA`), altering the injection locations: memory, CPU registers, and Program Counter (PC).

TABLE I: *Single number of faulty executions for each campaign and execution time for a single faulty execution.*

| Location | Faults | Injection Time (ms) |
|---|---|---|
| **Dijkstra** | | |
| Memory | 3330 | 260 |
| Registers | 5180 | 261 |
| PC | 155 | 30293 |
| **QSort** | | |
| Memory | 3595 | 257 |
| Registers | 5116 | 262 |
| PC | 161 | 30286 |
| **SHA** | | |
| Memory | 3677 | 254 |
| Registers | 15444 | 259 |
| PC | 168 | 30262 |

Table I reports the injection time for a single execution for all campaigns. These data show similar results for all the benchmarks and between memory and CPU registers campaigns. Instead, the time needed to complete a faulty execution in PC campaigns is very high compared to the others. This is due to the high percentage of crashes/hangs of these campaigns. The crash/hangs outcome is recognized by the FI script in the host leveraging a timeout exception, which takes time before occurring. For this reason, the number of faults generated in the PC campaigns is lower compared to the others. In general, to compute the total amount of time needed perform a single campaign, the number of generated faults (Table I), the number of executions needed for a single fault to collect all the architectural events, and the time taken by a single faulty execution must be considered (Table I).

The experiments targeted all used memory addresses for memory injections according to the ELF header, which reports all sections without distinguishing between OS and application sections. Regarding CPU registers, the solely standard registers were corrupted, excluding critical registers like the PC, which is specifically targeted in certain fault injection campaigns. During all experimental campaigns, the capability of profiling HPC was used for further analysis. To reduce the injection time, only architectural events that, from preliminary studies, exhibit changes and are not consistently constant were tracked. Eventually, every injected fault was labeled as Benign, SDC, or Other to characterize the fault effect.

Figure 4 illustrates injection outcomes based on fault locations and benchmarks. In CPU register campaigns, the predominant outcome is Benign (95%, 96.2%, and 88.4% for `Dijkstra`, `QSort`, and `SHA`). SDC is the second most common result, with percentages of 4.4%, 2.6%, and 11.2% for `Dijkstra`, `QSort`, and `SHA` respectively. Crashes/hangs are infrequent (0.6%, 1.2%, and 0.4% for `Dijkstra`, `QSort`, and `SHA`). The scarcity of crashes and hangs aligns with expectations, as special registers like Program Counter (PC) and Stack Pointer are excluded from these campaigns, focusing solely on standard registers. Notably, the elevated SDC instances related to the `SHA` execution during the CPU register campaign may be attributed to the increased CPU operations required for computing results over the same input data. Conversely, the higher occurrence of crashes and hangs in `QSort` and `Dijkstra` during CPU register campaigns, compared to `SHA`, suggests their generation from extensive use of structs and vectors in memory indexed by pointers, potentially leading to unexpected program behavior or crashes.

In PC injection campaigns, crashes/hangs dominate (89.1%, 83.9%, and 85.7% for `Dijkstra`, `QSort`, and `SHA`). Bit-flipping in PC causes significant flow modification. Corruption of the less significant byte leads to more heterogeneous outcomes, with SDC (1.9%, 0.6%, and 7.7%) and Benign (9%, 15.5%, and 6.6%). Specifically, corruption of the two least significant bits results in a benign outcome [30].

Memory injection campaigns show outcomes aligning with CPU register campaigns, with a strong majority of Benign results (96.9%, 98.5%, and 95% for `Dijkstra`, `QSort`, and `SHA`). SDCs occur in low percentages (0.7%, 0.9%, and 3.9%), attributed to the larger memory space compared to standard CPU registers, which also store data not currently used by the application at the injection instant. Crashes/hangs are infrequent (2.4%, 0.6%, and 1.1% for `Dijkstra`, `QSort`, and `SHA`).

Figure 3a highlights the profiling capabilities of the proposed fault injector. It reports a scatter plot of the fir two principal components of the `QSort` Principal Component Analysis (PCA), including all tracked features with outcomes labeled as Benign or SDC. The PCA is computed using architectural events as features, revealing the distribution of these two categories. The collected data were preprocessed via z-normalization, which normalizes every value in a dataset such that the mean of all values is 0 and the standard deviation is 1. Then Gaussianization is performed to apply transformation so that the data distribution of the transformed data is as Gaussian as possible. Generally, SDCs exhibit greater dispersion with a higher prevalence of outliers than the Benign category. This behavior is expected because bit flipping leading to an SDC likely introduces a nondeterministic behavior, causing a significant alteration of the architectural events. Looking at the Benign class, data are more concentrated, but variations still exist between executions. This observation suggests that soft errors can disrupt temporal constraints even if they do not alter the outcome. Further investigations are required to delve into this analysis. Nonetheless, the developed FI is valuable for such analyses.

In particular, Figure 3b shows the distribution of the Cycles (z-normalized and Gaussianized). The histogram plot highlights the differences concerning the number of cycles between the Benign and the SDCs, as SDCs present a higher deviation, which also means a higher time variability during the execution. However, benign distribution also exhibits variability, suggesting faults may introduce time deviations despite the correct final output, breaking the constraint of timing predictability demanded by SACRES.

## IV. CONCLUSION

This paper presented a new FI environment tailored for SACRES, capable of injecting bit-flips in real embedded hardware boards by exploiting the debug unit in modern CPUs. In its current implementation, injection in both CPU register and memory is possible. Differently from previous works, apart from providing a powerful and flexible FI environment, the proposed work also focuses on providing a powerful profiling tool based on the PMU available in modern microprocessors to profile desired architectural events using HPCs and collect them at the end of the computation. These additional data are fundamental when analyzing resilience to faults in real-time systems when hazards are not solely related to the correctness of the computation but also to the timing predictability of the system. Experimental results based on a Xilinx PYNQ Z2 board running FreeRTOS showed the capability and flexibility of the proposed framework. To encourage research in this field, we release the code related to our experiments as open-source: https://github.com/smilies-polito/marvin.

## ACKNOWLEDGMENT

This study was carried out within the "COLTRANE-V" project – funded by the Ministero dell'Università e della Ricerca – within the PRIN 2022 program (D.D.104 - 02/02/2022). This manuscript reflects only the authors' views and opinions, and the Ministry cannot be considered responsible for them.

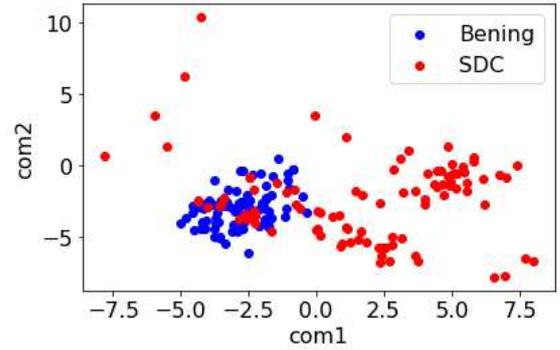

(a) *Scatter plot of the PCA-2 of all the features (architectural events).*

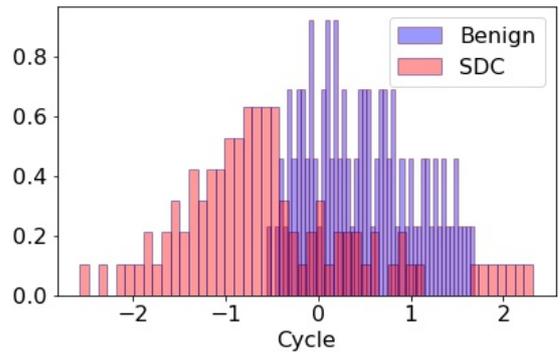

(b) *Histogram plot of the number of cycles distributions.*

Fig. 3: *Data visualization of QuickSort CPU registers campaigns dataset for the outcomes Benign and SDC. These figures highlight the different distributions of the two outcomes.*

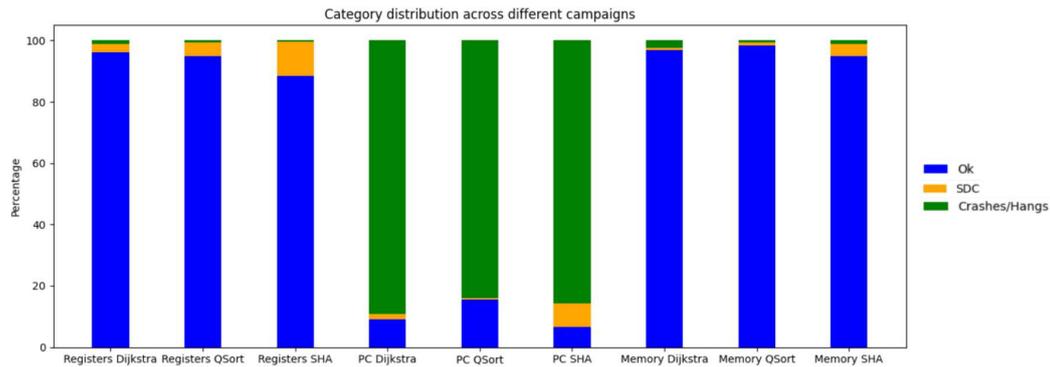

Fig. 4: *Breakdown of all injection campaigns. For each benchmark, classification percentages are reported for all potential targets: Registers, PC, and Memory.*